\documentclass[Physsubmission, Phys]{SciPost}

\hypersetup{
    colorlinks,
    linkcolor={red!50!black},
    citecolor={blue!50!black},
    urlcolor={blue!80!black}
}

\usepackage[bitstream-charter]{mathdesign}
\DeclareSymbolFont{usualmathcal}{OMS}{cmsy}{m}{n}
\DeclareSymbolFontAlphabet{\mathcal}{usualmathcal}

\newcommand{\ilog}{I_{log}(\lambda^{2})}
\newcommand{\iLog}{I_{log}^{(2)}(\lambda^{2})}
\newcommand{\ilogtwo}{I_{log}^{2}(\lambda^{2})}

\newcommand{\ti}[1]{\tilde{#1}}
\newcommand{\beq}{\begin{equation}}
\newcommand{\eeq}{\end{equation}}
\newcommand{\bea}{\begin{eqnarray}}
\newcommand{\eea}{\end{eqnarray} }

\begin{document}

\begin{center}{\Large \textbf{
Two-loop gauge coupling $\beta$-function  in a four-dimensional framework: the Standard Model case
}}\end{center}

\begin{center}
Adriano Cherchiglia\textsuperscript{1}
\end{center}

\begin{center}
{\bf 1} Centro de Ci\^ecias Naturais e Humanas, Universidade Federal do ABC, Santo Andr\'e, Brazil
\\
* adriano.cherchiglia@ufabc.edu.br
\end{center}

\begin{center}
\today
\end{center}


\definecolor{palegray}{gray}{0.95}
\begin{center}
\colorbox{palegray}{
  \begin{tabular}{rr}
  \begin{minipage}{0.1\textwidth}
    \includegraphics[width=35mm]{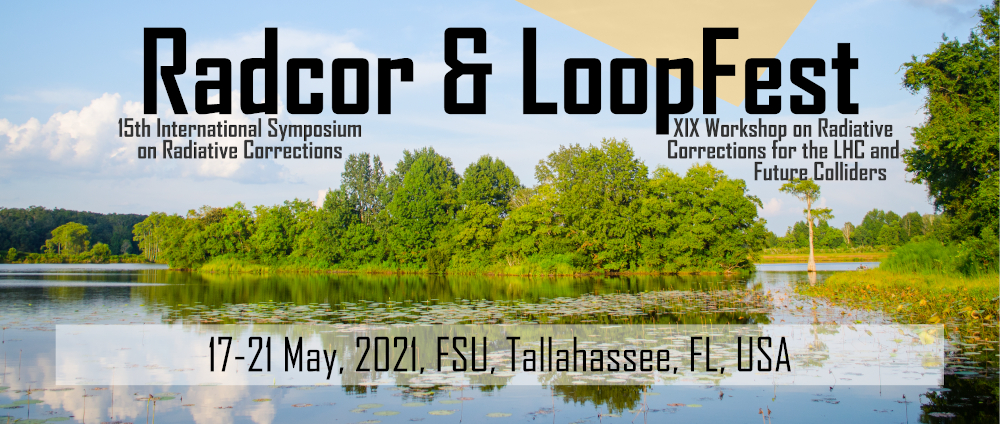}
  \end{minipage}
  &
  \begin{minipage}{0.85\textwidth}
    \begin{center}
    {\it 15th International Symposium on Radiative Corrections: \\Applications of Quantum Field Theory to Phenomenology,}\\
    {\it FSU, Tallahasse, FL, USA, 17-21 May 2021} \\
    \end{center}
  \end{minipage}
\end{tabular}
}
\end{center}

\section*{Abstract}
{\bf
At present, the gauge coupling $\beta$-function in the Standard Model (SM) is known up to four-loop order. As most SM calculations, dimensional regularization was employed. Despite its striking success, other regularization schemes have emerged, which aim to stay in the physical dimension as far as possible. In this contribution, we apply Implicit Regularization, a scheme defined in momentum space that complies with BPHZ, to obtain the two-loop gauge coupling $\beta$-function in the Standard Model. We reproduce the same result obtained when applying dimensional regularization.
}

\vspace{10pt}
\noindent\rule{\textwidth}{1pt}
\tableofcontents\thispagestyle{fancy}
\noindent\rule{\textwidth}{1pt}
\vspace{10pt}

\section{Introduction}
\label{sec:intro}

Renormalization group functions are essential ingredients to precisely probe the Standard Model and its extensions. They allow experimental results performed at different scales to be matched to theoretical predictions stemming from the knowledge of \textbf{divergent} intermediate pieces. Given the level of accuracy and accordance among experiment and theory, they can actually be seen as a triumph of the renormalization program implemented in the context of quantum field theory. The current state-of-the-art comprises the knowledge of the gauge coupling $\beta$-function in the SM at four-loop order \cite{Davies:2019onf,Bednyakov:2021qxa}, an achievement only possible after some recent developments regarding the treatment of the $\gamma_5$ matrix \cite{Poole:2019txl}. As customary, such calculations have been performed by relying on dimensional regularization, which may suffer from inconsistencies when dealing with dimensional specific objects. 

As an effort to circumvent these problems, other regularization methods have been proposed, which aim to stay in the physical dimension as far as possible (see \cite{Gnendiger:2017pys,TorresBobadilla:2020ekr} for a review). However, as recently pointed out \cite{Bruque:2018bmy}, they may also suffer from similar issues in general (see also \cite{Viglioni:2016nqc,Porto:2017asd,Batista:2018zxf} for particular examples). Nevertheless, since some of these methods aim to perform regularization/renormalization at integrand level, they may be more effective and prone to a numerical approach. One promising candidate is Implicit Regularization \cite{Battistel:1998sz} which is defined in momentum space and complies with BPHZ \cite{Cherchiglia:2010yd}. In order to sketch the  potential of this method, some working examples can be envisaged, one of those being the determination of the gauge coupling $\beta$-function in the SM at two-loop order. This constitutes a good check since the coefficients up to this loop order are universal as long as the subtraction schemes employed are mass-independent as is the case here. Moreover, this calculation may serve as a backbone to a more complete study of the RGE functions, which are regularization dependent at higher loop order in general, allowing the establishment of transition rules among methods. Given these prospects, in this contribution we obtain the gauge coupling $\beta$-function in the SM at two-loop order in the context of Implicit Regularization, confirming that it complies with known results.

We organize our work as follows: in section \ref{sec:ireg} we present a very brief overview of the IREG formalism, while in section \ref{sec:EW} we present our main results. Finally, we conclude in section \ref{sec:conclusion}.

\section{The Implicit Regularization formalism}
\label{sec:ireg} 

The Implicit Regularization (IREG) method, first presented in \cite{Battistel:1998sz}, is based on the idea that the UV divergent part of a general Feynman diagram can be isolated in integrals void from physical parameters (masses/external momenta). This is accomplished by applying a mathematical identity at integrand level which allows a separation between a UV only divergent integral (free of external momenta/masses) and a finite part. All in all, the UV divergent part of a general 1-loop Feynman amplitude will be proportional to
\bea
	I_{log}(\lambda^2)&\equiv& \int_{k} \frac{1}{(k^2-\lambda^2)^{2}},
	\eea
\noindent
where $\lambda$ plays the role of the renormalization group scale in the method.
By performing subtractions of $\ilog$ one defines a minimal subtraction scheme in the context of IREG. Therefore, the 1-loop renormalization functions $Z_{i}$ will be proportional to it as well. Since derivatives of $Z_{i}$ account for renormalization group functions, derivatives of $\ilog$ will be needed as well
\beq
\lambda^2\frac{\partial I_{log}(\lambda^2)}{\partial \lambda^2}= -b, \quad \quad b=\frac{i}{(4 \pi)^2}.
\label{eq:derivative}
\eeq

At higher loop order a similar program can be envisaged in a way compatible with Bogoliubov's recursion formula \cite{Cherchiglia:2010yd}. Since we are adopting a mass-independent subtraction scheme, the UV divergent part will only contain a logarithmic dependence, allowing us to define
\begin{align}
I_{log}^{(l)}(\lambda^2)&\equiv \int\limits_{k_{l}} \frac{1}{(k_{l}^2-\lambda^2)^{n}}
\ln^{l-1}{\left(-\frac{k_{l}^2-\lambda^2}{\lambda^2}\right)},
\end{align}
whose derivatives with respect to $\lambda$ can be straightforwardly obtained
\begin{align}
\lambda^2\frac{\partial I_{log}^{(l)}(\lambda^2)}{\partial \lambda^{2}}&=-(l-1)\, I_{log}^{(l-1)}(\lambda^2)- b \,\, (l-1)!\, \quad \mbox{where} \quad l>1.
\end{align}

Given these definitions, we can express the UV divergent part of a general Feynman amplitude at two-loop level as
\begin{align}
\mathcal{A}=c_{1}\;I_{log}^{(2)}(\lambda^2)+c_{2}\;\left[I_{log}(\lambda^2)\right]^{2}+c_{3}\;I_{log}(\lambda^2)
\end{align}
where $c_{i}$ are coefficients that may depend on masses/external momenta.

This brief description is aimed to provide the reader with the most basic and necessary ingredients of the IREG rules, easing the presentation of the results of upcoming sections. For further details we refer, for instance, to \cite{Arias-Perdomo:2021inz}.

\section{Gauge coupling $\beta$-function in the Standard Model to two-loop order}
\label{sec:EW} 

As a non-trivial example of the application of IREG we will pursue the computation of the two-loop coefficients for the gauge couplings $\beta$-function in the Standard Model. This improves the analysis performed in \cite {Cherchiglia:2020iug} where the $\beta$-function for an abelian (QED) and non-abelian (QCD) theory was studied.

\begin{figure}[h!]
\centering
\includegraphics[]{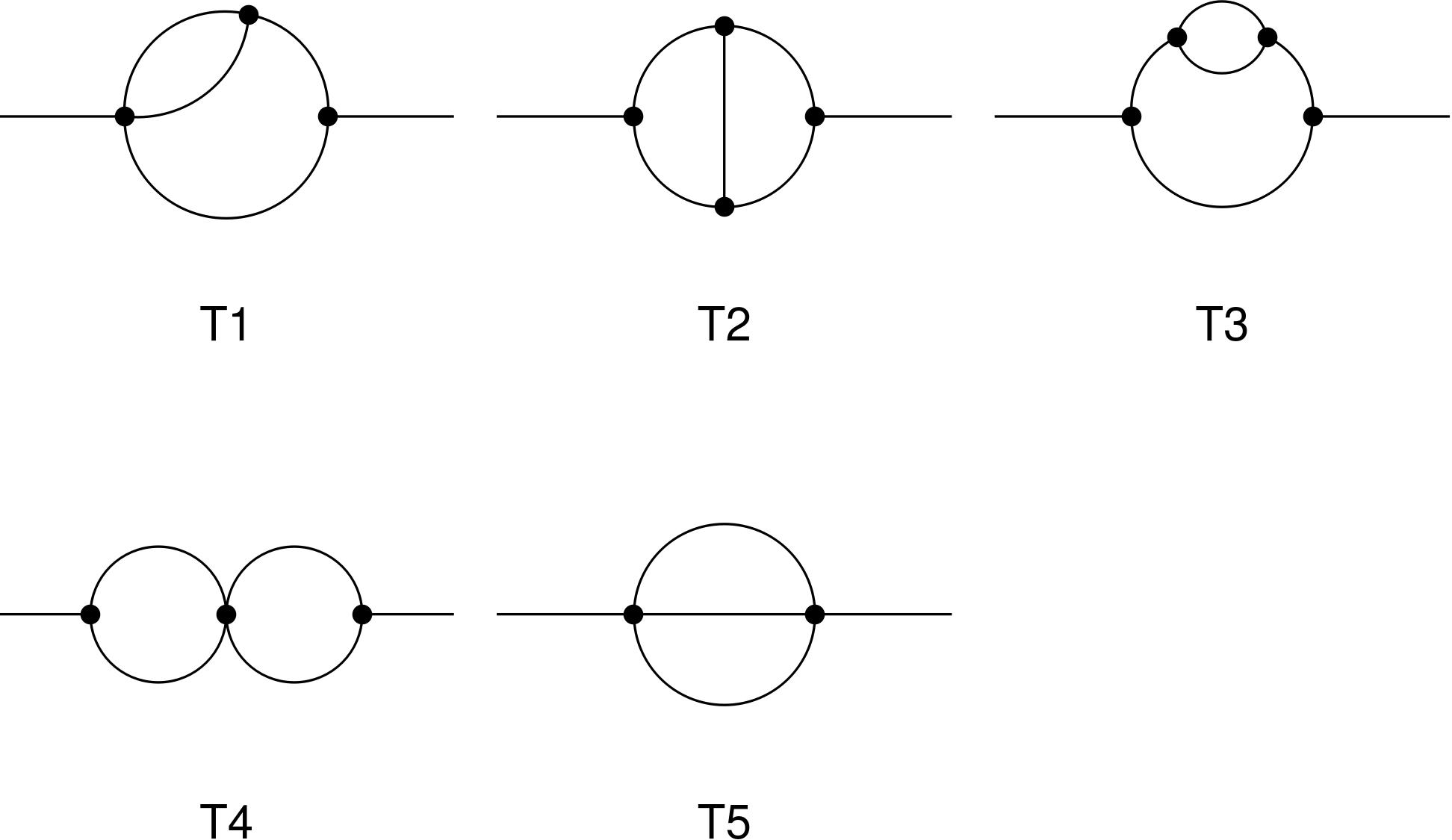}%
\caption{Two-loop topologies for two-point functions}
\label{Feynman}
\end{figure}

For this end, we will adopt the conventions of \cite{Mihaila:2012pz} and use the background field method \cite{Abbott:1980hw,Denner:1994xt}. This allows us to only compute two-point functions, instead of three-point ones needed in a standard computation. The generic topologies we are interested at can be seen in fig. \ref{Feynman}. 

\noindent
whose amplitudes are of the form
\begin{align}
\mathcal{A_{\text{T1}}} &\propto \int_{k,l} \frac{\mathcal{F}_{T1}(l,k,p)}{k^2(k-p)^2 l^2(l-k)^2},\nonumber\\
\mathcal{A_{\text{T2}}} &\propto \int_{k,l} \frac{\mathcal{F}_{T2}(l,k,p)}{k^2(k-p)^2 (k-l)^2 l^2(l-p)^2},\nonumber\\
\mathcal{A_{\text{T3}}} &\propto \int_{k,l} \frac{\mathcal{F}_{T3}(l,k,p)}{k^4(k-p)^2 l^2(l-k)^2},\nonumber\\
\mathcal{A_{\text{T4}}} &\propto \int_{k,l} \frac{\mathcal{F}_{T4}(l,k,p)}{k^2(k-p)^2 l^2(l-p)^2},\nonumber\\
\mathcal{A_{\text{T5}}} &\propto \int_{k,l} \frac{\mathcal{F}_{T5}(l,k,p)}{k^2 l^2 (l-k+p)^2}.
\end{align}   
Notice that we are considering massless propagators since we will only be interested in the UV divergent part of each amplitude.

The generic topologies will be filled with the particle content of the Standard Model by using the model file already implemented in \textit{FeynArts} \cite{Hahn:2000kx} supplemented with an extension related to the QCD background field method. 
A drawback of this approach is that only the theory on the spontaneously broken phase is available. Therefore, we will actually compute the 2-loop corrections to $\ti{A}\ti{A}$, $\ti{A}\ti{Z}$, $\ti{Z}\ti{Z}$, $\ti{G}\ti{G}$ where $\ti{A}$, $\ti{Z}$, $\ti{G}$ are the photon, Z-boson, and gluon background fields. The connection to the $\ti{B}$ and $\ti{W}$ fields is straightforward \cite{Mihaila:2012pz}
\begin{align}
\Pi_{\ti{B}\ti{B}}=\cos^{2}\theta_{W}^{\text{bare}}\Pi_{\ti{A}\ti{A}}+2\sin\theta_{W}^{\text{bare}}\cos\theta_{W}^{\text{bare}}\Pi_{\ti{A}\ti{Z}}+\sin^{2}\theta_{W}^{\text{bare}}\Pi_{\ti{Z}\ti{Z}}\\
\Pi_{\ti{W}\ti{W}}=\cos^{2}\theta_{W}^{\text{bare}}\Pi_{\ti{A}\ti{A}}-2\sin\theta_{W}^{\text{bare}}\cos\theta_{W}^{\text{bare}}\Pi_{\ti{A}\ti{Z}}+\sin^{2}\theta_{W}^{\text{bare}}\Pi_{\ti{Z}\ti{Z}}
\end{align}
As a check, we have explictly obtained  $\ti{W}\ti{W}$ as well, which is identical to the value obtained from the knowledge of $\Pi_{\ti{A}\ti{A}}$, $\Pi_{\ti{A}\ti{Z}}$, $\Pi_{\ti{Z}\ti{Z}}$. To perform the computation, we made use of \textit{FormCalc} \cite{Hahn:1998yk} as well as \textit{Package-X} \cite{Patel:2016fam}, which we embedded in our own routines. 
Since the number of diagrams is substantial, we refrain to present individual results, and we just quote the end result for the renormalization function of the background fields obtained in the context of IREG
\begin{align}
Z_{\ti{B}\ti{B}} =\; &1 - \frac{\alpha_{1}}{4\pi}\left[\frac{1}{10}+\frac{4}{3}n_{f}\right]\frac{\ilog}{b} + \frac{\alpha_{1}}{(4\pi)^{2}}\left[-\frac{9\alpha_{1}}{50}-\frac{9\alpha_{2}}{10}+\frac{17\text{tr}\hat{T}}{10}+\frac{\text{tr}\hat{B}}{2}+\frac{3\text{tr}\hat{L}}{2}\right.\nonumber\\
&\left.\quad-n_{f}\left(\frac{19\alpha_{1}}{5}+\frac{9\alpha_{2}}{5}+\frac{44\alpha_{3}}{15}\right)\right]\frac{\ilog}{b}\label{eq:za}\\
Z_{\ti{W}\ti{W}} =\; &1 - \frac{\alpha_{2}}{4\pi}\left[-\frac{43}{6}+\frac{4}{3}n_{f}\right]\frac{\ilog}{b} + \frac{\alpha_{2}}{(4\pi)^{2}}\left[-\frac{3\alpha_{1}}{10}\frac{\ilog}{b}\right.\nonumber\\
&\left.\quad+\frac{\alpha_{2}}{6b^2}\left(144(\ilogtwo-2b\iLog)+547b\ilog\right)\right.\nonumber\\
&\left.\quad+\left(\frac{3\text{tr}\hat{T}}{2}+\frac{3\text{tr}\hat{B}}{2}+\frac{\text{tr}\hat{L}}{2}\right)\frac{\ilog}{b}-n_{f}\left(\frac{3\alpha_{1}}{5}+49\alpha_{2}+4\alpha_{3}\right)\frac{\ilog}{b}\right]\label{eq:zb}\\
Z_{\ti{G}\ti{G}} =\; &1 - \frac{\alpha_{3}}{4\pi}\left[-11+\frac{4}{3}n_{f}\right]\frac{\ilog}{b} + \frac{\alpha_{3}}{(4\pi)^{2}}\left[\frac{6\alpha_{3}}{b^{2}}\left(9(\ilogtwo-2b\iLog)+35b\ilog\right)\right.\nonumber\\
&\left.+\left(2\text{tr}\hat{T}+2\text{tr}\hat{B}\right)\frac{\ilog}{b}-n_{f}\left(\frac{11\alpha_{1}}{10}+\frac{9\alpha_{2}}{2}+\frac{38\alpha_{3}}{3}\right)\frac{\ilog}{b}\right]
\label{eq:zc}
\end{align}
We have adopted the following conventions when writing our results\cite{Mihaila:2012pz}
\begin{align}
\alpha_{1}=\frac{5}{3}\frac{\alpha_{\text{QED}}}{\cos^{2}\theta_{W}}, \quad \alpha_{2}=\frac{\alpha_{\text{QED}}}{\sin^{2}\theta_{W}}, \quad \alpha_{3}=\alpha_{s},\quad \text{tr}\hat{X} = \frac{\alpha_{\text{QED}}}{2\sin^{2}\theta_{W}M_{W}^{2}}\sum_{i=1}^{n_{f}}m_{Xi}^{2},
\end{align}
where $\alpha_{\text{QED}}$ is the fine structure constant, $\theta_{W}$ is the Weinberg angle, $M_{W}$ is the W-boson mass, $\alpha_{s}$ is the QCD strong coupling constant, $m_{Xi}$ is the mass of the $X$-type fermion from the $i$-th generation, and $n_{f}$ is the number of generations.
We have also applied DREG and DRED, whose results agree with previous calculations \cite{Mihaila:2012pz}. Nevertheless, for comparison with IREG, we quote it here as well
\begin{align}
Z_{\ti{B}\ti{B}} =&\; 1 - \frac{\alpha_{1}}{4\pi}\left[\frac{1}{10}+\frac{4}{3}n_{f}\right]\frac{1}{\epsilon}\nonumber\\
& + \frac{\alpha_{1}}{(4\pi)^{2}}\left[-\frac{9\alpha_{1}}{100}-\frac{9\alpha_{2}}{20}+\frac{17\text{tr}\hat{T}}{20}+\frac{\text{tr}\hat{B}}{4}+\frac{3\text{tr}\hat{L}}{4}-n_{f}\left(\frac{19\alpha_{1}}{10}+\frac{9\alpha_{2}}{10}+\frac{22\alpha_{3}}{15}\right)\right]\frac{1}{\epsilon}\\
Z_{\ti{W}\ti{W}} =&\; 1 - \frac{\alpha_{2}}{4\pi}\left[-\frac{43}{6}+\frac{4}{3}n_{f}\right]\frac{1}{\epsilon} \nonumber\\
&+ \frac{\alpha_{2}}{(4\pi)^{2}}\left[-\frac{3\alpha_{1}}{20}+\frac{259\alpha_{2}}{12}+\frac{3\text{tr}\hat{T}}{4}+\frac{3\text{tr}\hat{B}}{4}+\frac{\text{tr}\hat{L}}{4}-n_{f}\left(\frac{3\alpha_{1}}{10}+\frac{49\alpha_{2}}{2}+2\alpha_{3}\right)\right]\frac{1}{\epsilon}\\
Z_{\ti{G}\ti{G}} =&\; 1 - \frac{\alpha_{3}}{4\pi}\left[-11+\frac{4}{3}n_{f}\right]\frac{1}{\epsilon}\nonumber\\
&+ \frac{\alpha_{3}}{(4\pi)^{2}}\left[51\alpha_{3}+\text{tr}\hat{T}+\text{tr}\hat{B}-n_{f}\left(\frac{11\alpha_{1}}{20}+\frac{9\alpha_{2}}{4}+\frac{17\alpha_{3}}{3}\right)\right]\frac{1}{\epsilon}
\end{align}

Some comments are in order: 1) when extracting the renormalization constants, we obtained that the end result is transverse, illustrating that IREG complies with non-abelian gauge invariance (see \cite{Cherchiglia:2020iug} for a more detailed analysis); 2) DREG and DRED differs in intermediate terms, but the final result is identical to each other as it should be (both subtraction schemes DR and MS are mass-independent \cite {Espriu:1981eh}); 3) counterterms for quantum fields are not needed as first discussed in \cite {Abbott:1980hw}, however, since we adopted the Feynman gauge, the renormalization for the gauge fixing was required.

Regarding the treatment of chiral fermions, we have applied two procedures available within \textit{FormCalc}. The first was just the naive scheme for $\gamma_{5}$, which implies that the $\gamma_{5}$ matrix could be anticommuted freely. 
In the second procedure we replaced $\gamma_{5}$ by its definition as
\beq
\gamma_{5}=-\frac{i}{4!}\epsilon_{abcd}{\gamma}^{a}{\gamma}^{b}{\gamma}^{c}{\gamma}^{d},
\label{eq:def G5}
\eeq
performed Dirac algebra together with contractions of the Levi-Civitá symbols that appear. Given the increased number of Dirac matrices, the second procedure demands more computational power. Both procedures amount to the same end results. We emphasize that none of the procedures is completely consistent in general. For the first procedure this is clear; for the second one the reason boils down to the use of identities involving Levi-Civitá symbols which in the context of IREG generates terms like $k^{2}$. However, since symmetric integration is not allowed in general in the method, such terms should be written as ${g}_{ab} k^{a}k^{b}$ instead which stands for enforcing the regularization before performing Lorentz contractions. For a more detailed discussion of these points we refer the reader to \cite{Bruque:2018bmy,Cherchiglia:2021uce}.
Finally, we would like to mention that ambiguities in the gauge couplings $\beta$-function in the Standard Model may arise only at four-loop level, at least in the context of dimensional methods \cite{Bednyakov:2015ooa}. Since the issues that IREG must face in the presence of $\gamma_{5}$ are similar to the ones encountered in dimensional methods, we also expect that only at higher loop order (potential) ambiguities may arise. 

To conclude this section, we present the gauge coupling $\beta$-function up to two-loop order obtained within the IREG framework. We adopt the same conventions of \cite{Mihaila:2012pz}
\begin{align}
\beta_{i}=\lambda^{2}\frac{d}{d \lambda^{2}} \frac{\alpha_{i}}{\pi}=-\frac{\alpha_{i}}{\pi}\lambda^{2}\frac{d}{d \lambda^{2}} \ln Z_{\alpha_{i}}=\frac{\alpha_{i}}{\pi}\lambda^{2}\frac{d}{d \lambda^{2}} \ln Z_{X_{i}}  
\end{align}
where $X_{i}$ denotes the background field corresponding to the coupling $\alpha_{i}$. Moreover, we can write
\begin{align}
Z_{X_{i}}&=1+\frac{\alpha_{i}}{4\pi}A_{i}+\sum_{j}\frac{\alpha_{i}\alpha_{j}}{(4\pi)^{2}}A_{ij}\\
\beta_{i}&=\left(\frac{\alpha_{i}}{4\pi}\right)^{2}\left[\beta_{i}+\sum_{j}\frac{\alpha_{j}}{4\pi}\beta_{ij}\right]
\end{align}
It is straightforward to obtain the relations, valid for IREG~\cite{Cherchiglia:2020iug},
\begin{align}
\beta_{i}=4\lambda^{2}\frac{d}{d \lambda^{2}}A_{i};\quad  \beta_{ij}=4\lambda^{2}\frac{d}{d \lambda^{2}}A_{ij}
\end{align}
Finally, using Eqs.~\ref{eq:za}-\ref{eq:zc}, yields
\begin{align}
\beta_{1} &= \frac{\alpha_{1}^{2}}{(4\pi)^{2}}\left[\frac{2}{5}+\frac{16}{3}n_{f}\right] + \frac{\alpha_{1}^{2}}{(4\pi)^{3}}\left[\frac{18\alpha_{1}}{25}+\frac{18\alpha_{2}}{5}-\frac{34\text{tr}\hat{T}}{5}-2\text{tr}\hat{B}-6\text{tr}\hat{L}\right.\nonumber\\
& \quad\quad\quad\quad\quad\quad\quad\quad\quad\quad\quad\quad\quad\quad\left.+n_{f}\left(\frac{76\alpha_{1}}{15}+\frac{12\alpha_{2}}{5}+\frac{176\alpha_{3}}{15}\right)\right]\\
\beta_{2} &= \frac{\alpha_{2}^{2}}{(4\pi)^{2}}\left[-\frac{86}{3}+\frac{16}{3}n_{f}\right] + \frac{\alpha_{2}^{2}}{(4\pi)^{3}}\left[\frac{6\alpha_{1}}{5}-\frac{518\alpha_{2}}{3}-6\text{tr}\hat{T}-6\text{tr}\hat{B}-2\text{tr}\hat{L}\right.\nonumber\\
& \quad\quad\quad\quad\quad\quad\quad\quad\quad\quad\quad\quad\quad\quad\left.+n_{f}\left(\frac{4\alpha_{1}}{5}+\frac{196\alpha_{2}}{3}+16\alpha_{3}\right)\right]\\
\beta_{3} &= \frac{\alpha_{3}^{2}}{(4\pi)^{2}}\left[-44+\frac{16}{3}n_{f}\right] + \frac{\alpha_{3}^{2}}{(4\pi)^{3}}\left[-408 \alpha_{3}-8\text{tr}\hat{T}-8\text{tr}\hat{B}\right.\nonumber\\
& \quad\quad\quad\quad\quad\quad\quad\quad\quad\quad\quad\quad\quad\quad\left.+n_{f}\left(\frac{22\alpha_{1}}{15}+6\alpha_{2}+\frac{304\alpha_{3}}{3}\right)\right]
\end{align}

As can be seen, we recover known results obtained in the context of dimensional methods in a minimal subtraction scheme \cite{Mihaila:2012pz}. This was expected since in both cases a subtraction scheme independent of the mass was used, which implies that up to two-loop order the gauge coupling $\beta$-function is renormalization scheme independent \cite{Espriu:1981eh}.

\section{Conclusion}
\label{sec:conclusion}

In the last years we have witnessed an unprecedented amount of data collected in the LHC. Since no clear sign of Beyond Standard Model (BSM) Physics has emerged, it is  required to know the Standard Model predictions with great precision to clear disentangle the possible BSM contributions. In order to tackle this task, innovative techniques have emerged, some of those proposing distinct regularization methods which aim to stay in the physical dimension as far as possible. Among those, Implicit Regularization is a promising candidate, aiming to perform regularization at integrand level. It has been shown to comply with unitarity, Lorentz invariance, causality, and abelian gauge symmetry in general, while it complies with non-abelian gauge invariance in working examples. In this contribution, we apply the technique for the first time in the Standard Model, which is a spontaneously broken chiral non-abelian theory. We have obtained the two-loop gauge coupling $\beta$-function, reproducing the results known when applying dimensional regularizations. This proves as a consistent test of the technique, since up to two-loop level the gauge coupling $\beta$-function is universal to any mass-independent subtraction scheme.



\paragraph{Funding information}
The author acknowledges support from Fundação para a Ciência e Tecnologia (FCT) through the project CERN/FIS-COM/0035/2019, and the networking support by the COST Action CA16201.  


%





\nolinenumbers

\end{document}